\documentclass[sigconf]{acmart}
\AtBeginDocument{%
  }

\setcopyright{acmlicensed}
\copyrightyear{2026}
\acmYear{2026}
\acmDOI{XXXXXXX.XXXXXXX}
\acmISBN{978-1-4503-XXXX-X/2018/06}

\usepackage{multirow}
\usepackage{xcolor}            
\usepackage{colortbl}          
\newcommand{\method}[1]{\textsc{#1}}
\newcommand{\model}{\method{\textbf{LLM4GTA}}{}}

\begin{document}

\title{Can Representation Gaps Be the Key to Enhancing Robustness in Graph-Text Alignment?}
\author{Heng Zhang}
\affiliation{
  \institution{South China Normal University}
  \country{China}}
\email{2024025450@m.scnu.edu.cn}

\author{Tianyi Zhang}
\affiliation{
  \institution{Uber Technologies Inc.}
  \country{USA}}
\email{tianyizhg@gmail.com}

\author{Yuling Shi}
\affiliation{
  \institution{Shanghai Jiao Tong University}
  \city{Shanghai}
  \country{China}}
\email{yuling.shi@sjtu.edu.cn}

\author{Xiaodong Gu}
\affiliation{
  \institution{Shanghai Jiao Tong University}
  \city{Shanghai}
  \country{China}}
\email{xiaodong.gu@sjtu.edu.cn}

\author{Yaomin Shen}
\affiliation{
  \institution{Nanchang Research Institute, Zhejiang University}
  \city{Nanchang}
  \country{China}}
\email{coolshennf@gmail.com}

\author{Zijian Zhang}
\affiliation{
  \institution{University of Pennsylvania}
  \country{USA}}
\email{zzjharry@alumni.upenn.edu}

\author{Yilei Yuan}
\affiliation{
  \institution{University of Michigan}
  \country{USA}}
\email{yiliey@umich.edu}

\author{Hao Zhang}
\affiliation{
  \institution{University of Chinese Academy of Sciences}
  \country{China}}
\email{zh.cs.star@outlook.com}

\author{Jin Huang}
\authornote{Corresponding author.}
\affiliation{
  \institution{South China Normal University}
  \country{China}}
\email{huangjin@m.scnu.edu.cn}

\renewcommand{\shortauthors}{Trovato et al.}

\begin{abstract}
Representation learning on text-attributed graphs (TAGs) integrates structural connectivity with rich textual semantics, enabling applications in diverse domains. Current methods largely rely on contrastive learning to maximize cross-modal similarity, assuming tighter coupling between graph and text representations improves transfer performance. However, our empirical analysis reveals that both natural gap expansion and forced gap reduction result in performance degradation by disrupting pre-trained knowledge structures and impairing generalization. This arises from the geometric incompatibility between encoders, where graph encoders capture topological patterns, while text encoders capture semantic structures. Over-alignment compresses these distinct spaces into shared subspaces, causing structure collapse that diminishes both topological reasoning and semantic understanding. We propose \textbf{LLM4GTA}, a gap-aware alignment framework that preserves representation gaps as geometric necessities for maintaining modality-specific knowledge and improving transfer performance. LLM4GTA includes an adaptive gap preservation module to prevent over-alignment by monitoring similarity evolution and an intra-modal compensation mechanism that boosts discriminative power using auxiliary classifiers in graph space. Extensive experiments show significant improvements over existing methods in zero-shot and few-shot scenarios.
\end{abstract}

\ccsdesc[500]{Information systems~Data mining}
\keywords{Graph Foundation Model, Graph Representation Learning, Few-shot Learning} 


\maketitle

\section{Introduction}
Text-attributed graphs (TAGs) are an effective way to model complex relationships between textual entities in various real-world applications \cite{gpt4graph,unigraph}, including social networks \cite{GraphEdit}, knowledge bases \cite{feng2024kgt}, and academic citation networks \cite{graphclip}. Unlike traditional graphs, TAGs enrich nodes with textual information, providing additional semantic context \cite{graphprompt}. For instance, in academic citation networks, nodes represent research papers with attributes like titles or abstracts, and edges signify citation relationships. Recent research focuses on aligning graph structures with textual information, aiming to create unified representations that better capture the interplay between structural and semantic features, thereby enabling more accurate analysis and reasoning \cite{ENGINE,graph2text}.

\begin{figure}[tb!]
    \centering
    \includegraphics[width=\columnwidth]{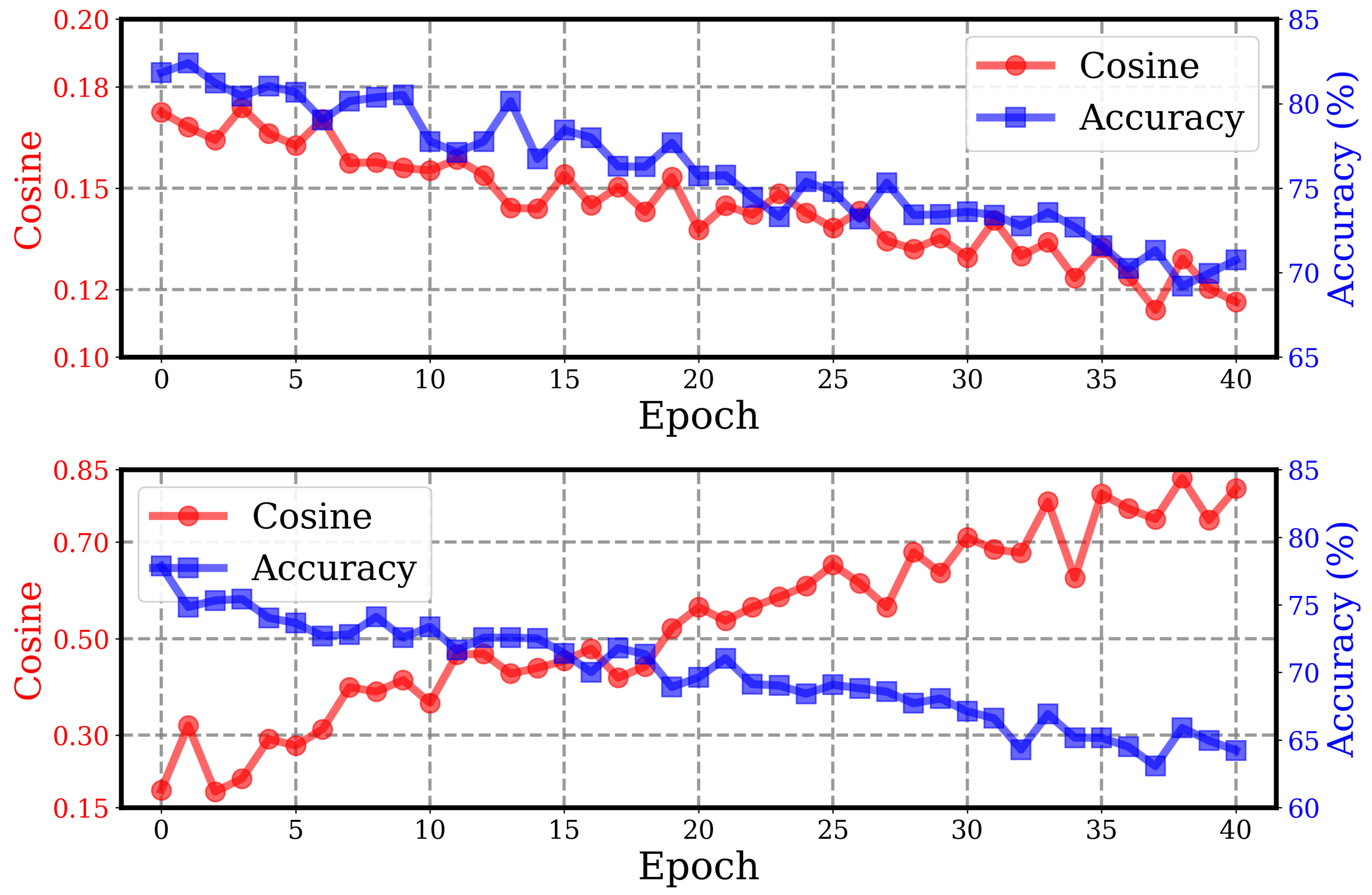}
    \caption{Representation similarity and accuracy evolution on PubMed. Top: Naive task-specific fine-tuning causes representation drift, with both similarity and accuracy declining. Bottom: Contrastive alignment loss increases graph-text similarity but struggle to prevent accuracy degradation. 
    }
    \vspace{-0.15in}
    \label{figure1}
\end{figure}
Research on graph-text alignment has largely progressed along two key directions. The first focuses on contrastive learning frameworks, aiming to map graph and text representations into a shared embedding space by maximizing their mutual information \cite{wang2024llms,natureisneed,huang2023can}. Representative methods, such as GraphCLIP \cite{graphclip} and ConGrat \cite{congrat}, employ InfoNCE loss to align graph and text representations, typically freezing pre-trained text encoders while adapting graph encoders to fit the text representation space. This strategy treats alignment as a process of bridging the gap between graph structures and textual semantics. The second direction centers on unified encoder architectures that directly integrate graph and text processing through advanced fusion mechanisms \cite{GCA,NLGraph,mvgrl}. For example, approaches like ENGINE \cite{ENGINE} and LLaGA \cite{chen2024llaga}, and CTGL \cite{zhou2025ctgl} incorporate graph neural networks and language models with cross-attention layers to capture interactions between structural and semantic information \cite{zhou2025ctgl}. Additionally, recent works such as GAugLLM \cite{fang2024gaugllm} and STAG~\cite{stag2025kdd} leverage large language models to augment graph-text data, improving the quality of contrastive training samples \cite{GraphEdit,graphinsight}. Despite their differences, these methods share a common goal: enhancing alignment between graph and text features to achieve better performance on downstream tasks.

To understand the relationship between alignment strength and model performance, we conducted empirical studies tracking representation changes during graph-text contrastive training. Current methods assume tighter coupling between graph and text representations leads to better transfer learning. This assumption drives the design of contrastive objectives that maximize similarity between matched pairs. We empirically investigate this assumption by monitoring both cosine similarity and accuracy on PubMed throughout the training process under two contrasting scenarios (Figure~\ref{figure1}). Our experiments reveal an unexpected pattern. As shown in Figure~\ref{figure1}(top), naive task-specific fine-tuning with standard cross-entropy loss causes the representation gap to expand uncontrollably. Cosine similarity drops from 0.18 to 0.12 while accuracy declines from 82\% to 68\%. The expanding gap accompanies knowledge forgetting as the two representation spaces naturally drift apart. Figure~\ref{figure1}(bottom) presents a more puzzling scenario. Introducing contrastive alignment loss successfully narrows the representation gap, increasing similarity from 0.15 to over 0.70. Yet accuracy still deteriorates throughout training, dropping from 78\% to 62\%. This decoupling between similarity and performance challenges the core assumption. Neither allowing natural gap expansion nor forcing gap reduction produces superior results. Both extremes disrupt the balanced state established during pre-training. These observations prompts us to reconsider: \textbf{might the gap between graph and text encoders serve a similar protective role rather than representing an alignment failure?}

The experimental findings expose a deeper issue in current graph-text alignment approaches. Graph encoders learn through message passing mechanisms in graph neural networks. They encode topological patterns through iterative neighborhood aggregation. The learned representations capture structural properties like homophily patterns. These representations inhabit a geometric space shaped by graph topology. Text encoders learn through self-attention mechanisms in language models. They encode semantic relations through contextual associations. The learned representations capture linguistic patterns. These representations inhabit a geometric space shaped by language distribution. The two encoding processes create representations on geometrically distinct manifolds. They differ in intrinsic dimensionality, curvature properties, and metric structure. 

Contrastive alignment forces these distinct spaces into a shared subspace. This compression causes structure collapse in two dimensions: \textbf{\textit{(i) Topological signals in graph representations get smoothed out.}} The features distinguishing different structural patterns become compressed. Graph homophily patterns break down. This resembles over-smoothing in graph neural networks but occurs across representation spaces. Graph encoders lose the structural inductive biases enabling generalization to new topologies. \textbf{\textit{(ii) The semantic space of text representations gets distorted.}} Semantic relations stretch to match graph geometry. The compositional structure learned during pretraining deteriorates. Language model patterns lose integrity. Text encoders lose the semantic prior knowledge acquired from pretraining. From an information-theoretic view, graph encoders learn mutual information between representations and topology. Text encoders learn mutual information between representations and semantics. Over-maximizing their mutual alignment inevitably reduces both quantities. This destroys the domain-invariant structural and semantic knowledge essential for transfer learning.

Building on the previous analysis, we introduce the LLM4GTA framework. Unlike traditional methods that attempt to eliminate modality gaps, this framework recognizes gaps as essential for preserving the specialized knowledge of each encoder. Appropriate separation safeguards the geometric integrity of pretrained representations. This is particularly important when downstream data lacks the diversity needed to represent the full range of pretraining distributions. Forcing alignment based on such limited data can distort global structures and weaken pretrained knowledge crucial for transfer learning. The representation gap provides two critical benefits. It preserves structural inductive biases from graph encoders, enabling generalization to new topologies, and retains semantic prior knowledge from language models, supporting zero-shot reasoning with unseen concepts. To address these challenges, LLM4GTA introduces two core solutions. The first is an adaptive gap preservation module that dynamically monitors representation distances during training. It tracks positive pair similarities to ensure meaningful correspondences and negative pair similarities to maintain space separation. When harmful convergence patterns are detected, the module triggers early stopping to avoid structural collapse while supporting healthy adaptation. The second is an intra-graph-space compensation mechanism. Auxiliary classifiers trained directly in the graph representation space enhance discriminative power by fully leveraging structural features without being constrained by cross-space distances. Our main contributions are as follows:
\begin{itemize}
    \item We reveal that representation gaps protect encoder-specific geometric structures essential for cross-domain generalization in graph-text alignment. Maintaining appropriate separation enhances transfer performance.
    \item We propose \textbf{LLM4GTA}, a gap-aware graph-text alignment framework that preserves appropriate modality gaps to maintain pre-trained knowledge and enable effective cross-modal learning without extensive labeled data.
    \item Extensive experiments show that LLM4GTA outperforms state-of-the-art methods, achieving 8.24\% higher accuracy on citation networks and 6.91\% on social networks in both zero-shot and few-shot settings.
\end{itemize}

\begin{figure*}[t]
   \centering
   \includegraphics[width=\textwidth]{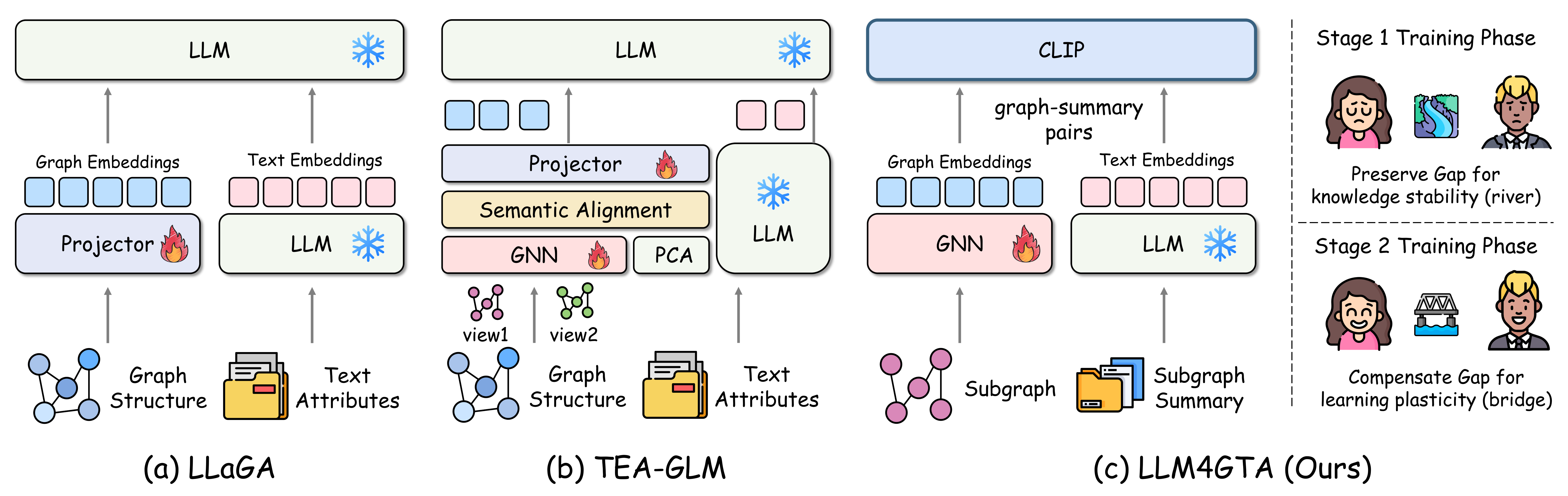}
\caption{Comparison of graph-text alignment methods for text-attributed graphs. 
(a) LLaGA employs a trainable projector to map GNN-encoded graph tokens into 
LLM embedding space. (b) TEA-GLM performs token-level alignment between GNN 
representations and LLM embeddings. (c) Our method builds upon self-supervised 
contrastive pretraining with LLM-generated summaries, introducing gap-aware 
adaptation through modality gap monitoring for training stability and 
dual-space classification for enhanced task adaptability.}
   \label{fig:main}
\end{figure*}

\section{Related Work}
\subsection{LLMs for Graphs}
The integration of LLMs into graph learning has yielded methods that often surpass traditional GNN approaches. Current research explores diverse integration strategies \cite{gpt4graph,unigraph,huang2023can}. Enhancement-based methods leverage LLMs to enrich node features with semantic information before feeding them to downstream models \cite{graph2text,NLGraph,llm2graph}. Direct prediction approaches serialize graph structures into natural language, enabling LLMs to perform end-to-end inference \cite{natureisneed,graphprompt,graphgpt,chen2024exploring,zhao2024gimlet}. Beyond these primary paradigms, LLMs serve various auxiliary roles \cite{yu2023empower,wang2023graph}. GraphEdit \cite{GraphEdit} employs them for data augmentation, RLMRec \cite{RLMRec} aligns GNN embeddings with linguistic knowledge, and ENG \cite{ENGINE} generates synthetic nodes to address data scarcity. Recent alignment-based methods like GraphClip \cite{graphclip} bridge GNN representations with LLM embeddings, achieving impressive zero-shot performance across diverse tasks and datasets \cite{GraphText,GraphTranslator,GraphWiz,gpt4graph}. Emerging hybrid architectures demonstrate the synergy between GNNs and LLMs, with frameworks combining structural graph embeddings with language model capabilities to enhance both retrieval and generation tasks \cite{mavromatis2024gnnrag,li2024hybridllmgnn}. The rise of Graph Retrieval-Augmented Generation (GraphRAG) has further advanced the field by leveraging knowledge graphs to improve contextual understanding and reduce hallucinations in LLM outputs \cite{msftgraphrag2024}. These approaches incorporate graph-based indexing, guided retrieval, and enhanced generation pipelines to capture relational knowledge more effectively than traditional vector-based RAG systems \cite{peng2024graphrag,han2025graphrag}. Knowledge graph-enhanced frameworks have also shown remarkable success in domain-specific applications, where structured relational information significantly improves reasoning accuracy and factual grounding \cite{feng2024kgt}.

\subsection{Graph Few-shot Learning}
Real-world graph applications often face limited labeled data, making few-shot and zero-shot learning crucial for practical deployment \cite{GCC,JOAO,graphgpt}. Early approaches leverage meta-learning paradigms to enable rapid adaptation with minimal examples \cite{zhu2020deep,GCA}. Methods such as Meta-GNN \cite{zhu2020deep}, G-Meta \cite{GCA}, and Meta-GPS \cite{GCC} rely on a bi-level optimization strategy based on MAML, aiming to achieve better parameter initialization. While these models deliver strong results, their reliance on diverse meta-training tasks and their high complexity have hindered further progress. Self-supervised methods, including DGI \cite{DGI} and GraphCL \cite{GraphCL}, construct contrastive views to enhance node representations through pre-training, whereas MVGRL \cite{mvgrl} incorporates subgraph and diffusion information to capture richer graph semantics. Similarly, TLP \cite{GCC} and TEG \cite{GCA} attempt to address the limited diversity of meta-training datasets using graph contrastive learning and equivariant neural networks, yielding promising results in few-shot scenarios. However, these methods typically require task-specific fine-tuning and may degrade significantly when supervision becomes extremely sparse. Recent advances explore pseudo-labeling, expanding limited labeled data through confident predictions on unlabeled nodes \cite{ye2023natural,chen2023label,he2023harnessing}. The emergence of LLMs provides new possibilities for addressing these challenges \cite{wang2024can}. For example, LLaGA \cite{chen2024llaga} introduces encoding strategies that enable LLMs to process graph data without training, while TEA-GLM \cite{tea-glm} aligns GNN representations with LLM embeddings to achieve state-of-the-art cross-task and cross-dataset generalization. These advancements demonstrate that LLMs can effectively tackle graph learning challenges without task-specific supervision \cite{GraphLLM,GraphInstruct,graphclip}.

\section{Preliminaries}
\subsection{Graph-Text Alignment Task Definition}
We consider an alignment learning task based on pre-trained graph-text models. Given a text-attributed graph $G = \{V, \{T_n\}_{n=1}^N, A\}$, where $V$ denotes the node set with $|V| = N$, $A \in \mathbb{R}^{N \times N}$ represents the adjacency matrix, and $T_n$ denotes the raw text attributes for node $n \in [1, 2, \ldots, N]$. Node features $X \in \mathbb{R}^{N \times d}$ are derived by applying sentence encoders to the raw text information, where $d$ represents the feature dimension. The objective of graph-text alignment is to learn mapping functions $f_g: \mathcal{G} \rightarrow \mathbb{R}^d$ and $f_t: \mathcal{T} \rightarrow \mathbb{R}^d$ that project graph structural information and textual semantic information into a unified representation space. Here, $\mathcal{G}$ denotes the space of graph structures and $\mathcal{T}$ represents the space of textual descriptions. The learned representations should preserve both the topological properties inherent in graph structures and the semantic relationships embedded in textual content. For scalability considerations, we employ subgraph sampling to extract manageable portions from large graphs. A sampling function $\Gamma(\cdot)$ is applied to derive a set of subgraphs $\mathcal{I} = \{G_n\}_{n=1}^N$, where each $G_n$ represents a subgraph centered on node $n$. This sampling strategy enables efficient processing of large-scale graph data while maintaining representative structural patterns. The graph-text alignment task aims to establish semantic correspondences between graph representations $\mathbf{h}_i = f_g(G_i)$ and text representations $\mathbf{t}_j = f_t(T_j)$, such that semantically related graph-text pairs exhibit high similarity while unrelated pairs remain distinguishable. This alignment enables cross-domain knowledge transfer and zero-shot learning capabilities on unseen graph-text combinations.

\subsection{Representation Gap Measurement for Graph-Text Learning}
In graph-text alignment tasks, the representation gap refers to the natural distance between graph and text representation spaces in the learned embedding space. This gap reflects the intrinsic differences between structural and semantic information processing mechanisms. Given $N$ graph nodes and $K$ class texts, we define the overall cross-representation similarity as the average cosine similarity between all graph node features and text features:
\begin{equation}
\text{sim}_{overall} = \frac{1}{N} \sum_{i=1}^{N} \frac{1}{K} \sum_{j=1}^{K} \cos(\mathbf{h}_i, \mathbf{t}_j),
\end{equation}
where $\mathbf{h}_i \in \mathbb{R}^d$ represents the representation vector of graph node $i$, $\mathbf{t}_j \in \mathbb{R}^d$ represents the representation vector of text class $j$, and $\cos(\cdot, \cdot)$ denotes the cosine similarity function. To analyze the impact of graph-text similarity in a more fine-grained manner, we decompose the overall similarity into positive and negative components. The average similarity of positive graph-text pairs is defined as:
\begin{equation}
\text{sim}_{pos} = \frac{1}{N}\sum_{i=1}^N\cos(\mathbf{h}_i, \mathbf{t}_{y_i}),
\end{equation}
where $\mathbf{t}_{y_i}$ denotes the text representation corresponding to the ground truth class $y_i$ of graph node $i$. For graphs with heterogeneous degree distributions, we can optionally weight nodes by their structural importance $w_i = (\text{deg}(i) + 1) / (\max_j \text{deg}(j) + 1)$ to emphasize well-connected nodes, though we use uniform weighting in our main experiments for simplicity. Similarly, we define the average similarity of negative graph-text pairs as:
\begin{equation}
\text{sim}_{neg} = \frac{1}{N} \sum_{i=1}^{N} \frac{1}{K-1} \sum_{j \neq y_i} \cos(\mathbf{h}_i, \mathbf{t}_j),
\end{equation}
where the inner summation covers all text representations $\mathbf{t}_j$ that do not correspond to the ground truth class of graph node $i$. The representation gap can be quantified as the difference between positive and negative similarities:
\begin{equation}
\text{gap} = \text{sim}_{pos} - \text{sim}_{neg}
\end{equation}
A well-calibrated representation gap should maintain $\text{sim}_{pos} > \text{sim}_{neg}$ while avoiding excessive separation that might indicate overfitting to specific domains. These measurements provide quantitative tools for monitoring representation quality throughout the learning process. By tracking changes in $\text{sim}_{pos}$, $\text{sim}_{neg}$, and their difference, we can identify when alignment procedures begin to compromise the stability of pre-trained representations.

\section{Methodology}
Our LLM4GTA addresses graph-text alignment from a representation gap perspective, proposing gap-aware strategies that balance knowledge preservation with learning plasticity. The framework comprises three components: gap analysis, gap preservation, and gap compensation. By monitoring representation changes during training, we maintain each encoder's original capabilities while achieving stable cross-modal learning. 

\subsection{Understanding Representation Gap Dynamics in Graph-Text Learning}
We first analyze how contrastive learning objectives affect representation gaps between graph and text encoders. This analysis reveals why naive alignment approaches often lead to degraded generalization performance.

\subsubsection{Gap Expansion During Contrastive Optimization}
The standard contrastive loss function for graph-text alignment is defined as:
\begin{equation}
\mathcal{L}_{CL} = -\log \frac{\exp(\text{sim}(\mathbf{h}_i, \mathbf{t}_{y_i}) / \tau)}{\sum_{j=1}^{K} \exp(\text{sim}(\mathbf{h}_i, \mathbf{t}_j) / \tau)},
\end{equation}
where $\mathbf{h}_i \in \mathbb{R}^d$ represents the representation vector of graph node $i$, $\mathbf{t}_j \in \mathbb{R}^d$ represents the text representation vector of class $j$, $\text{sim}(\cdot, \cdot)$ denotes the cosine similarity function, $\tau$ is the temperature parameter, and $y_i$ represents the ground truth label of node $i$. This optimization objective attempts to push positive pair similarity toward 1 and negative pairs toward 0, conflicting with the moderate similarity distribution observed in well-performing pre-trained models.The optimization process fundamentally alters the geometric relationship between graph and text representation spaces. In pre-trained models, these representations naturally cluster within distinct but semantically meaningful regions. However, contrastive learning forces these distinct clusters to collapse toward identical points, eliminating the representational diversity that enables robust cross-domain transfer.

\subsubsection{Structural Imbalance in Graph-Text Classification}
Unlike symmetric text-image pre-training scenarios, graph-text classification tasks exhibit significant structural imbalance. Each text class must establish positive relationships with all graph nodes in that class while maintaining negative relationships with all other classes. This "one-to-many" matching pattern results in negative samples far outnumbering positive samples, causing optimization processes to be dominated by negative sample repulsion operations. 

The mathematical formulation of this imbalance can be expressed as:
\begin{equation}
\mathbb{E}[\mathcal{L}_{neg}] = \frac{C-1}{C} \mathbb{E}[\log \sum_{j \neq y_i} \exp(\text{sim}(\mathbf{h}_i, \mathbf{t}_j) / \tau)],
\end{equation}
where $C$ represents the total number of classes and $\mathcal{L}_{neg}$ represents the loss contribution from negative samples. As $C$ increases, the negative sample contribution dominates the optimization, leading to excessive representation separation. This imbalance creates a cascading effect where the model increasingly focuses on distinguishing between different classes rather than learning meaningful semantic correspondences.

\subsubsection{GNN Over-Smoothing and Gap Collapse}
Graph neural networks aggregate neighborhood information through message passing, which can lead to over-smoothing when combined with aggressive alignment. When contrastive loss forces $\mathbf{h}_i \approx \mathbf{t}_{y_i}$ for all nodes in the same class, it causes within-class node representations to collapse toward a single point, losing structural diversity. We quantify this through intra-class variance $\text{Var}_c = \frac{1}{|V_c|}\sum_{i \in V_c} \|\mathbf{h}_i - \bar{\mathbf{h}}_c\|^2$, where $V_c$ denotes nodes in class $c$ and $\bar{\mathbf{h}}_c$ is the class centroid. Empirically, naive alignment reduces $\text{Var}_c$ by over 60\%, while our gap preservation maintains 85\% of the original variance, demonstrating that controlled gap maintenance prevents structural collapse in GNN representations.

\subsection{Representation Stability through Gap Preservation}
Based on our analysis of gap dynamics, we design a gap monitoring mechanism to maintain representation stability. This mechanism operates by tracking the evolution of positive and negative sample similarities throughout training.

\subsubsection{Training Phase Analysis}
We observe distinct phases during the training process. Early training primarily improves positive sample similarities, indicating meaningful semantic learning. However, later training phases begin degrading negative sample similarities, signaling the onset of representation space distortion. We define the gap change rate as:
\begin{equation}
\rho = \frac{d(\text{sim}_{pos} - \text{sim}_{neg})}{dt},
\end{equation}
where $\text{sim}_{pos}$ and $\text{sim}_{neg}$ represent positive and negative sample similarities respectively, and $t$ represents the training time step. When $\rho$ transitions from positive to negative values, it indicates excessive gap expansion is beginning.

\subsubsection{Early Stopping Mechanism}
We propose an early stopping mechanism based on negative sample similarity monitoring. The mechanism first establishes a baseline by computing the original model's negative sample similarity $\text{sim}_{neg}^{base}$ on a validation set. During training, we continuously monitor the current negative sample similarity $\text{sim}_{neg}^{curr}$. The relative change is quantified as:
\begin{equation}
\delta = \frac{|\text{sim}_{neg}^{curr} - \text{sim}_{neg}^{base}|}{\text{sim}_{neg}^{base}},
\end{equation}
where $\text{sim}_{neg}^{base}$ serves as the reference point representing the optimal representation gap state, and $\text{sim}_{neg}^{curr}$ captures the current state during training. When $\delta$ exceeds a preset threshold $\theta$, training is terminated to prevent excessive representation gap changes. In practice, we find that the optimal $\theta$ may vary with graph properties: denser graphs (higher average degree) can tolerate slightly larger gap changes, while sparse graphs require more conservative thresholds. For our experiments, we use $\theta = 0.10$ for citation networks and $\theta = 0.12$ for social networks, selected via validation on the first task. This mechanism ensures preservation of pre-trained knowledge structures by preventing the representation space distortion that occurs during aggressive alignment.

\subsection{Multi-Space Classification through Gap Compensation}
While gap preservation maintains representation stability, it may limit the model's discriminative capabilities. To address this limitation, we construct a compensatory classifier operating directly in the graph representation space.

\subsubsection{Theoretical Foundation for Graph Space Classification}
From a theoretical perspective, the optimal classifier $\mathbf{W}^*$ should exist within the span of the input feature space. For graph-text alignment problems, text classifiers are constrained by the dimensional limitations of the text representation space and cannot fully utilize discriminative information available in the graph representation space. The text representation matrix $\mathbf{T}$ can be decomposed as:
\begin{equation}
\mathbf{T} = \mathbf{T}_{\parallel} + \mathbf{T}_{\perp}
\end{equation}
where $\mathbf{T}_{\parallel}$ lies in the subspace spanned by graph features $\mathbf{H}$, and $\mathbf{T}_{\perp}$ is orthogonal to it. Since text features generally do not fully span the graph feature space, the best achievable text classifier is restricted to a lower-dimensional subspace, leading to suboptimal classification boundaries. 

We define a graph space classifier $\mathbf{W}_g \in \mathbb{R}^{d \times C}$ with the optimization objective:
\begin{equation}
\mathbf{W}_g^* = \arg\min_{\mathbf{W}_g} \mathbb{E}[\ell(\mathbf{W}_g^T \mathbf{h}_i, y_i)]
\end{equation}
where $\ell(\cdot, \cdot)$ represents the classification loss function, $\mathbf{h}_i$ represents node representations produced by the frozen graph encoder, $d$ is the representation dimension, and $C$ is the number of classes. This classifier directly learns discriminative boundaries in the graph representation space, circumventing the constraints imposed by cross-modal alignment requirements.

\subsubsection{Dual-Space Prediction Fusion}
The dual-space prediction fusion strategy combines cross-modal semantic understanding with uni-modal discriminative capabilities. The fusion mechanism operates by weighted combination of predictions from both representation spaces. The final prediction probability is computed as:
\begin{equation}
P(y|\mathbf{h}_i) = \text{softmax}(\text{sim}(\mathbf{h}_i, \mathbf{T}) + \lambda \cdot \mathbf{W}_g^T \mathbf{h}_i),
\end{equation}
where $\mathbf{T} = [\mathbf{t}_1, \mathbf{t}_2, \ldots, \mathbf{t}_C] \in \mathbb{R}^{d \times C}$ represents the text representation matrix containing all class representations, $\lambda \in \mathbb{R}^+$ is the weight parameter that balances the contributions of cross-modal and uni-modal predictions, and $\mathbf{h}_i \in \mathbb{R}^d$ is the graph representation of node $i$. The first term $\text{sim}(\mathbf{h}_i, \mathbf{T})$ leverages cross-modal alignment to provide semantic understanding capabilities, enabling the model to benefit from rich textual knowledge embedded in pre-trained language models. The second term $\lambda \cdot \mathbf{W}_g^T \mathbf{h}_i$ enhances discriminative capability through graph space classification, allowing the model to learn domain-specific decision boundaries without being constrained by representation gap preservation requirements. This fusion strategy ensures that the model maintains both semantic consistency across domains and discriminative precision within specific domains. The weight parameter $\lambda$ can be tuned to emphasize either cross-modal generalization or uni-modal discrimination based on the specific requirements of target applications.

\begin{table*}[tbp] 
    \centering
    \small
     \addtolength{\tabcolsep}{0.7mm}
    \caption{Accuracy (\%) of one-shot \emph{node classification} with standard deviations. Each column represents a target domain, using other columns as source domains.  The best method in each column is bolded, and the runner-up is underlined.
    }
    \label{table1}%
    \begin{tabular}{c|ccccccc}
    \toprule
   {{Method }\textbackslash{ Target domain}}   & Cora & Citeseer & Pubmed & Photo & Computers & Facebook & LastFM
      \\\midrule
    \method{GCN} 
    & 29.53 $\pm$ \phantom{0}7.56 
    & 26.29 $\pm$ \phantom{0}6.50  
    & 23.32 $\pm$ 11.56  
    & 26.96 $\pm$ 12.94 
    & 24.40 $\pm$ \phantom{0}5.62 
    & 20.45 $\pm$ \phantom{0}5.62 
    & \phantom{0}9.21 $\pm$ \phantom{0}3.11   
\\ 
    \method{GAT} 
    & 24.27 $\pm$ \phantom{0}9.26  
    & 21.56 $\pm$ \phantom{0}8.09    
    & 22.28 $\pm$ \phantom{0}9.76   
    & 17.85 $\pm$ 10.22 
    & 23.03 $\pm$ 12.12 
    & 29.27 $\pm$ \phantom{0}6.47   
    & \phantom{0}9.01 $\pm$ \phantom{0}2.61
 
\\\midrule
    \method{DGI}
    & 33.40 $\pm$ 10.48  
    & 25.80 $\pm$ \phantom{0}8.27
    & 47.22 $\pm$ \phantom{0}9.50  
    & 30.89 $\pm$ 10.54  
    & 25.75 $\pm$ 12.45  
    & 34.36 $\pm$ \phantom{0}9.57 
    & 14.14 $\pm$ \phantom{0}6.31
\\
    \method{GraphCL}
    & 27.72 $\pm$ \phantom{0}9.37   
    & 35.02 $\pm$ \phantom{0}8.46  
    & 48.89 $\pm$ \phantom{0}9.03  
    & 34.78 $\pm$ 11.56  
    & {23.79} $\pm$ 12.28 
    & 34.85 $\pm$  \phantom{0}7.07 
    & 18.93 $\pm$  \phantom{0}7.32 
    
\\
    \method{GPPT}
    & 27.18 $\pm$ \phantom{0}4.88	
    & 25.90 $\pm$ \phantom{0}4.68 
    & 39.82 $\pm$ \phantom{0}8.79 
    & 31.58 $\pm$ 10.27  
    & 19.94 $\pm$ \phantom{0}9.61
    & 34.73 $\pm$ \phantom{0}3.99 
    & 20.98 $\pm$ \phantom{0}3.98
\\
    \method{GraphPrompt}
    & 28.26 $\pm$ 12.68
    & 32.51 $\pm$ \phantom{0}8.73
    & 47.47 $\pm$ \phantom{0}9.15
    & 48.11 $\pm$ \phantom{0}9.89  
    & 42.82 $\pm$ 11.67 
    & 40.44 $\pm$ \phantom{0}9.68
    & 19.84 $\pm$ \phantom{0}7.23 
\\
    \method{LLaGA}
    & 31.93 $\pm$ 10.24
    & 33.17 $\pm$ 11.85
    & 48.36 $\pm$ \phantom{0}7.92
    & 51.28 $\pm$ 10.43  
    & 43.67 $\pm$ \phantom{0}9.18 
    & 41.89 $\pm$ 10.77
    & 22.13 $\pm$ \phantom{0}8.46 
\\
    \method{GraphGPT}
    & 29.45 $\pm$ \phantom{0}8.73
    & 35.24 $\pm$ \phantom{0}9.31
    & 49.82 $\pm$ 10.68
    & 49.76 $\pm$ \phantom{0}8.54  
    & 45.31 $\pm$ 12.29 
    & 42.15 $\pm$ \phantom{0}7.93
    & 20.98 $\pm$ \phantom{0}6.85
\\
    \method{TEA-GLM}
    & 30.72 $\pm$ 13.45
    & 34.89 $\pm$ \phantom{0}7.26
    & 48.93 $\pm$ \phantom{0}8.41
    & 50.45 $\pm$ 11.72  
    & 44.58 $\pm$ 10.83 
    & 41.37 $\pm$ \phantom{0}8.25
    & 21.76 $\pm$ \phantom{0}9.12
\\
    \method{GraphInsight}
    & 28.84 $\pm$ \phantom{0}9.18
    & 33.76 $\pm$ 10.42
    & 49.15 $\pm$ \phantom{0}9.87
    & 52.33 $\pm$ \phantom{0}9.16  
    & 43.95 $\pm$ \phantom{0}8.91 
    & 42.28 $\pm$ 11.34
    & 23.47 $\pm$ \phantom{0}7.68
\\
    \method{GraphTranslator}
    & 32.18 $\pm$ 11.76
    & 34.45 $\pm$ \phantom{0}8.97
    & \underline{50.21} $\pm$ 10.25
    & 49.87 $\pm$ 10.61  
    & 46.13 $\pm$ \phantom{0}9.54 
    & 41.72 $\pm$ \phantom{0}9.41
    & 21.35 $\pm$ \phantom{0}8.93
\\
    \method{GPF}
    & 32.17 $\pm$ \phantom{0}6.56
    & 36.79 $\pm$ \phantom{0}7.70 
    & 41.28 $\pm$ \phantom{0}8.14 
    & 47.47 $\pm$ \phantom{0}8.19  
    & 35.75 $\pm$ \phantom{0}7.12 
    & 40.45 $\pm$ \phantom{0}6.34
    & 27.26 $\pm$ \phantom{0}5.50
\\\midrule
    \method{Hassani}
    & 33.35 $\pm$ \phantom{0}6.93
    & 33.66 $\pm$ \phantom{0}7.24
    & 39.87 $\pm$ \phantom{0}8.16
    & 48.48 $\pm$ \phantom{0}7.07
    & 39.99 $\pm$ \phantom{0}7.91
    & 37.70 $\pm$ \phantom{0}5.79
    & 27.16 $\pm$ \phantom{0}4.94
    
\\\midrule

    \method{GCOPE}
    & \underline{35.62} $\pm$ 11.93	
    & \textbf{38.33} $\pm$ \phantom{0}9.28  
    & 45.38 $\pm$ \phantom{0}9.87  
    & \underline{52.87} $\pm$ \phantom{0}9.19
    & \underline{48.65} $\pm$ 10.69 
    & \underline{43.63} $\pm$ \phantom{0}8.50
    & \underline{28.84} $\pm$ \phantom{0}7.59
 \\
    \rowcolor[HTML]{E1EAFF}
      \method{\model}
    & \textbf{52.37} $\pm$ 12.64  
    & \underline{38.18} $\pm$ \phantom{0}9.75
    & \textbf{54.76} $\pm$ 11.28 
    & \textbf{62.94} $\pm$ \phantom{0}9.83 
    & \textbf{51.33} $\pm$ \phantom{0}8.92 
    & \textbf{47.18} $\pm$ \phantom{0}9.41
    & \textbf{36.29} $\pm$ \phantom{0}8.76  
    
\\    \bottomrule
        \end{tabular}
        \\
\end{table*}

\begin{table*}[tbp] 
    \centering
    \small
     \addtolength{\tabcolsep}{0.7mm}
    \caption{Accuracy (\%) of one-shot \emph{graph classification} with standard deviations. Each column represents a target domain, using other columns as source domains.  The best method in each column is bolded, and the runner-up is underlined.
    }
    \label{table2}%
    \begin{tabular}{c|ccccccc}
    \toprule
   {Method }\textbackslash{ Target domain}   & Cora & Citeseer & Pubmed & Photo & Computers & Facebook & LastFM
      \\\midrule
     \method{GCN} 
    & 30.64 $\pm$ 10.31 
    & 26.90 $\pm$ \phantom{0}7.15 
    & 38.84 $\pm$ 11.82 
    & 15.60 $\pm$ \phantom{0}8.77   
    & 21.94 $\pm$ 14.51 
    & 31.33 $\pm$ \phantom{0}9.47  
    & 28.83 $\pm$ \phantom{0}9.60
    
\\ 
    \method{GAT} 
    & 27.80 $\pm$ \phantom{0}7.85  
    & 27.50 $\pm$ \phantom{0}7.13    
    & 21.66 $\pm$ \phantom{0}8.70    
    & 15.74 $\pm$ \phantom{0}7.62 
    & 16.02 $\pm$ 13.46 
    & 21.20 $\pm$ \phantom{0}7.31  
    & 27.80 $\pm$ \phantom{0}7.85
 
\\\midrule
    \method{InfoGraph}
    & 34.98 $\pm$ 10.15 
    & 35.87 $\pm$ \phantom{0}9.84
    & 48.67 $\pm$ 12.29  
    & 25.70 $\pm$ 11.73  
    & 19.02 $\pm$ 14.09  
    & 31.26 $\pm$ \phantom{0}9.65 
    & 23.29 $\pm$ \phantom{0}7.99
\\
    \method{GraphCL}
    & 42.70 $\pm$ 10.64
    & 36.66 $\pm$ \phantom{0}8.67
    & 47.53 $\pm$ 11.52  
    & 33.07 $\pm$ 12.31  
    & 16.02 $\pm$ 13.47 
    & 21.99 $\pm$ 13.00
    & 21.30 $\pm$ 10.45
\\
    \method{GraphPrompt}
    & 37.38 $\pm$ 14.03	
    & 36.66 $\pm$ \phantom{0}9.19  
    & 49.55 $\pm$ 10.25 
    & 50.79 $\pm$ 12.31
    & 43.09 $\pm$ 11.45 
    & 41.71 $\pm$ 10.61
    & 32.62 $\pm$ \phantom{0}8.54
\\
    \method{LLaGA}
    & 40.73 $\pm$ 12.87	
    & 37.28 $\pm$ 10.42  
    & 48.91 $\pm$ 11.63 
    & 52.44 $\pm$ 13.05
    & 45.18 $\pm$ 10.29 
    & 40.85 $\pm$ 11.94
    & 35.27 $\pm$ \phantom{0}9.18
\\
    \method{GraphEdit}
    & 38.56 $\pm$ 11.74	
    & 39.12 $\pm$ \phantom{0}8.95  
    & 47.83 $\pm$ 12.47 
    & 51.62 $\pm$ 11.89
    & 46.34 $\pm$ 12.73 
    & 39.96 $\pm$ \phantom{0}9.82
    & 33.95 $\pm$ 10.36
\\
    \method{TEA-GLM}
    & 41.29 $\pm$ 13.56	
    & 35.84 $\pm$ 10.73  
    & 48.42 $\pm$ \phantom{0}9.88 
    & 53.16 $\pm$ 10.47
    & 44.87 $\pm$ 13.15 
    & 41.03 $\pm$ 10.28
    & 36.48 $\pm$ \phantom{0}8.91
\\
    \method{GraphInsight}
    & 39.15 $\pm$ 14.62	
    & 38.47 $\pm$ \phantom{0}9.81  
    & 49.26 $\pm$ 11.94 
    & 50.93 $\pm$ 12.76
    & \underline{47.52} $\pm$ 11.89 
    & 40.42 $\pm$ 12.35
    & 34.19 $\pm$ \phantom{0}7.82
\\
    \method{GraphTranslator}
    & 38.92 $\pm$ 12.19	
    & 37.95 $\pm$ 11.36  
    & 48.17 $\pm$ 10.72 
    & 52.88 $\pm$ 11.24
    & 45.61 $\pm$ 10.96 
    & 41.38 $\pm$ 11.57
    & 33.74 $\pm$ \phantom{0}9.43
\\
    \method{GPF}
    & 39.62 $\pm$ \phantom{0}8.52	
    & 36.73 $\pm$ \phantom{0}7.66 
    & 45.08 $\pm$ 10.36 
    & 47.57 $\pm$ 10.16 
    & 35.70 $\pm$ \phantom{0}8.71  
    & 34.84 $\pm$ \phantom{0}5.14  
    & 34.31 $\pm$ \phantom{0}7.05

\\\midrule
    \method{Hassani}
    & 36.86 $\pm$ 10.74
    & 35.78 $\pm$ \phantom{0}8.80
    & 43.97 $\pm$ 13.27
    & 41.55 $\pm$ 13.08
    & 29.49 $\pm$ 13.86
    & 35.57 $\pm$ \phantom{0}9.00
    & 25.39 $\pm$ \phantom{0}8.14
\\\midrule

    \method{GCOPE}
    & \underline{43.85} $\pm$ 10.99	
    & \textbf{42.93} $\pm$ \phantom{0}9.82 
    & \underline{51.05} $\pm$ 11.74
    & \underline{56.93} $\pm$ \phantom{0}9.74  
    & 45.60 $\pm$ 10.96 
    & \underline{43.26} $\pm$ \phantom{0}9.53
    & \underline{37.68} $\pm$ \phantom{0}7.70
 \\
  \rowcolor[HTML]{E1EAFF}
      \method{\model}
    & \textbf{59.87} $\pm$ 14.31  
    & \underline{40.64} $\pm$ 10.17
    & \textbf{52.48} $\pm$ 10.89 
    & \textbf{63.16} $\pm$ 12.25 
    & \textbf{52.94} $\pm$ 11.76 
    & \textbf{46.82} $\pm$ 10.13
    & \textbf{51.75} $\pm$ 10.38 
\\    \bottomrule
        \end{tabular}
\end{table*}

\begin{table*}[tbp] 
    \centering
    \small
     \addtolength{\tabcolsep}{0.7mm}
    \caption{Accuracy (\%) of \underline{zero-shot} \emph{node classification} with standard deviations. Each column represents a target domain, using other columns as source domains.  The best method in each column is bolded, and the runner-up is underlined.
    }
    \label{table3}%
    \begin{tabular}{c|ccccccc}
    \toprule
   {{Method }\textbackslash{ Target domain}}   & Cora & Citeseer & Pubmed & Photo & Computers & Facebook & LastFM
      \\\midrule
    \method{GCN} 
    & 27.84 $\pm$ \phantom{0}8.23 
    & 24.75 $\pm$ \phantom{0}7.18  
    & 21.89 $\pm$ 12.34  
    & 25.13 $\pm$ 13.67 
    & 22.91 $\pm$ \phantom{0}6.28 
    & 19.12 $\pm$ \phantom{0}6.15 
    & \phantom{0}8.47 $\pm$ \phantom{0}3.56   
\\ 
    \method{GAT} 
    & 22.93 $\pm$ 10.14  
    & 20.28 $\pm$ \phantom{0}8.76    
    & 20.84 $\pm$ 10.42   
    & 16.52 $\pm$ 11.08 
    & 21.48 $\pm$ 13.05 
    & 27.65 $\pm$ \phantom{0}7.23   
    & \phantom{0}8.31 $\pm$ \phantom{0}2.94
 
\\\midrule
    \method{DGI}
    & 31.47 $\pm$ 11.23  
    & 24.28 $\pm$ \phantom{0}9.01
    & 44.36 $\pm$ 10.18  
    & 28.93 $\pm$ 11.37  
    & 24.12 $\pm$ 13.28  
    & 32.47 $\pm$ 10.34 
    & 13.21 $\pm$ \phantom{0}6.89
\\
    \method{GraphCL}
    & 26.18 $\pm$ 10.05   
    & 32.84 $\pm$ \phantom{0}9.27  
    & 45.92 $\pm$ \phantom{0}9.78  
    & 32.67 $\pm$ 12.34  
    & 22.34 $\pm$ 13.15 
    & 32.81 $\pm$  \phantom{0}7.84 
    & 17.76 $\pm$  \phantom{0}8.01 
    
\\
    \method{GPPT}
    & 25.63 $\pm$ \phantom{0}5.42	
    & 24.37 $\pm$ \phantom{0}5.29 
    & 37.45 $\pm$ \phantom{0}9.46 
    & 29.73 $\pm$ 11.08  
    & 18.73 $\pm$ 10.38
    & 32.68 $\pm$ \phantom{0}4.56 
    & 19.72 $\pm$ \phantom{0}4.43
\\
    \method{GraphPrompt}
    & 26.72 $\pm$ 13.52
    & 30.57 $\pm$ \phantom{0}9.48
    & 44.63 $\pm$ 10.07
    & 45.28 $\pm$ 10.67  
    & 40.26 $\pm$ 12.43 
    & 38.12 $\pm$ 10.52
    & 18.61 $\pm$ \phantom{0}7.89 
\\
    \method{LLaGA}
    & 30.18 $\pm$ 11.07
    & 31.25 $\pm$ 12.63
    & 45.51 $\pm$ \phantom{0}8.56
    & 48.37 $\pm$ 11.18  
    & 41.08 $\pm$ \phantom{0}9.87 
    & 39.47 $\pm$ 11.52
    & 20.83 $\pm$ \phantom{0}9.14 
\\
    \method{GraphGPT}
    & 27.84 $\pm$ \phantom{0}9.41
    & 33.18 $\pm$ 10.08
    & 46.92 $\pm$ 11.45
    & 46.84 $\pm$ \phantom{0}9.27  
    & 42.67 $\pm$ 13.12 
    & 39.73 $\pm$ \phantom{0}8.67
    & 19.73 $\pm$ \phantom{0}7.48
\\
    \method{TEA-GLM}
    & 29.03 $\pm$ 14.28
    & 32.84 $\pm$ \phantom{0}7.93
    & 46.08 $\pm$ \phantom{0}9.15
    & 47.56 $\pm$ 12.48  
    & 41.97 $\pm$ 11.59 
    & 38.98 $\pm$ \phantom{0}9.02
    & 20.47 $\pm$ \phantom{0}9.87
\\
    \method{GraphInsight}
    & 27.23 $\pm$ \phantom{0}9.93
    & 31.78 $\pm$ 11.19
    & 46.28 $\pm$ 10.61
    & 49.28 $\pm$ \phantom{0}9.89  
    & 41.37 $\pm$ \phantom{0}9.64 
    & 39.85 $\pm$ 12.13
    & 22.08 $\pm$ \phantom{0}8.34
\\
    \method{GraphTranslator}
    & 30.43 $\pm$ 12.58
    & 32.45 $\pm$ \phantom{0}9.71
    & 47.29 $\pm$ 10.98
    & 46.98 $\pm$ 11.37  
    & 43.47 $\pm$ 10.28 
    & 39.32 $\pm$ 10.17
    & 20.08 $\pm$ \phantom{0}9.68
\\
    \method{GPF}
    & 30.42 $\pm$ \phantom{0}7.18
    & 34.63 $\pm$ \phantom{0}8.37 
    & 38.84 $\pm$ \phantom{0}8.87 
    & 44.73 $\pm$ \phantom{0}8.92  
    & 33.67 $\pm$ \phantom{0}7.78 
    & 38.13 $\pm$ \phantom{0}6.97
    & 25.63 $\pm$ \phantom{0}6.07
\\\midrule
    \method{Hassani}
    & 31.52 $\pm$ \phantom{0}7.58
    & 31.68 $\pm$ \phantom{0}7.89
    & 37.53 $\pm$ \phantom{0}8.83
    & 45.67 $\pm$ \phantom{0}7.74
    & 37.64 $\pm$ \phantom{0}8.56
    & 35.52 $\pm$ \phantom{0}6.38
    & 25.57 $\pm$ \phantom{0}5.48
    
\\\midrule

    \method{GCOPE}
    & \underline{33.72} $\pm$ 12.78	
    & \underline{36.18} $\pm$ \phantom{0}9.97  
    & \underline{47.85} $\pm$ 10.52  
    & \underline{50.12} $\pm$ \phantom{0}9.94
    & \underline{45.87} $\pm$ 11.42 
    & \underline{41.18} $\pm$ \phantom{0}9.23
    & \underline{27.13} $\pm$ \phantom{0}8.26
 \\
    \rowcolor[HTML]{E1EAFF}
      \method{\model}
    & \textbf{49.78} $\pm$ 13.42  
    & \textbf{36.23} $\pm$ 10.48
    & \textbf{51.94} $\pm$ 12.05 
    & \textbf{59.86} $\pm$ 10.57 
    & \textbf{48.76} $\pm$ \phantom{0}9.68 
    & \textbf{44.82} $\pm$ 10.18
    & \textbf{34.53} $\pm$ \phantom{0}9.47  
    
\\    \bottomrule
        \end{tabular}
        \\
\end{table*}


\section{Experiments}
\subsection{Datasets}
We evaluate LLM4GTA on seven diverse text-attributed graph datasets spanning multiple domains to assess cross-domain transferability. The target datasets include citation networks with Cora~\cite{sen2008collective}, Citeseer~\cite{sen2008collective}, and Pubmed~\cite{sen2008collective} representing academic publications, e-commerce graphs with Photo~\cite{shchur2018pitfalls} and Computers~\cite{shchur2018pitfalls} capturing product relationships, and social networks with Facebook~\cite{traud2012social} and LastFM~\cite{rozemberczki2020characteristic} modeling user interactions. Each dataset exhibits distinct structural properties and semantic characteristics, enabling comprehensive evaluation of transfer learning capabilities across heterogeneous graph types. Following standard protocols, we train on large-scale source datasets and evaluate zero-shot and few-shot performance on these target domains.

\subsection{Baselines}
We compare LLM4GTA against fifteen baseline methods across three categories. Traditional graph learning methods include GCN \cite{kipf2017semi} and GAT \cite{velivckovic2018graph} for supervised learning, along with self-supervised approaches DGI and GraphCL that leverage contrastive objectives. Graph prompting methods comprise GPPT \cite{sun2023gppt} and GraphPrompt \cite{graphprompt} designed for few-shot adaptation. Recent LLM-based methods include LLaGA \cite{chen2024llaga} and GraphGPT \cite{graphgpt} that integrate language models with graph encoders \cite{GraphEdit}, TEA-GLM \cite{wang2024llms} and GraphInsight \cite{graphinsight} that perform token-level alignment, and GraphTranslator that bridges graph and text representations. Additional competitive baselines include GPF \cite{xia2024gpf} for graph foundation models, Hassani's cross-domain method \cite{hassani2020contrastive}, and GCOPE \cite{gcope2024} for graph-text co-training. All methods follow identical evaluation protocols with the same train-test splits to ensure fair comparison across zero-shot and few-shot scenarios.
\begin{figure}[t]
   \centering
   \includegraphics[width=\columnwidth]{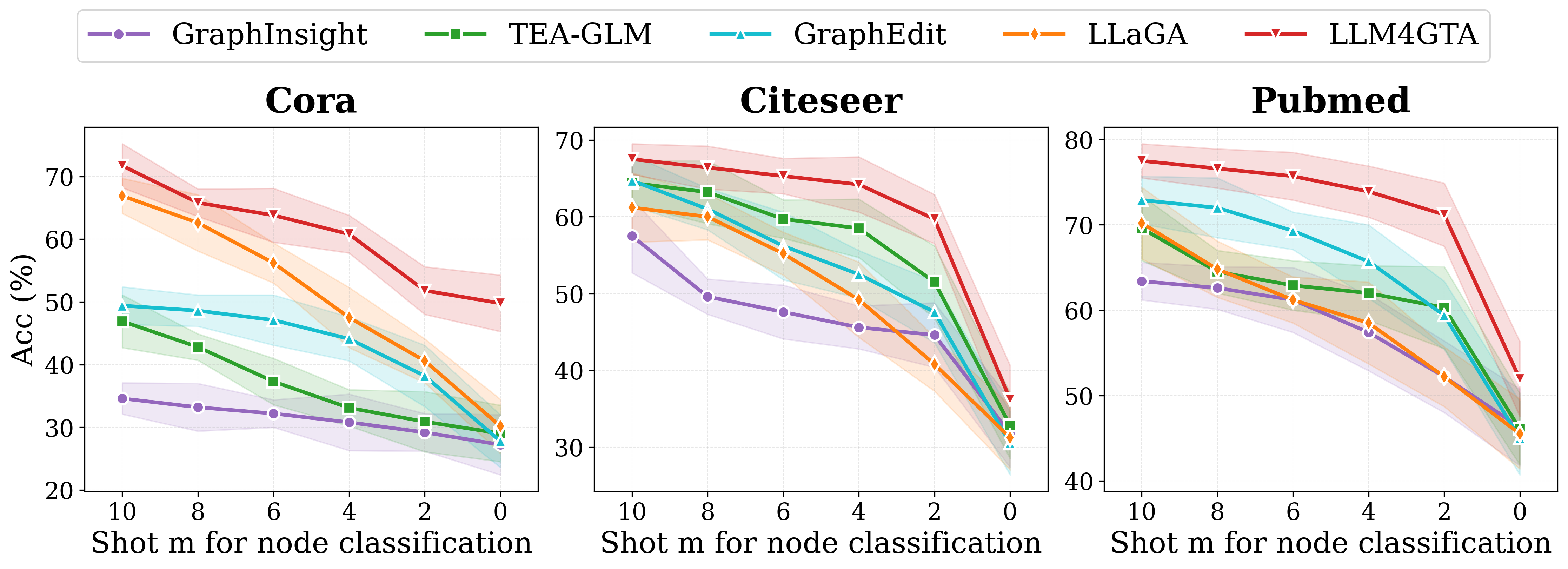}
\caption{Impact of number of shots on node classification on three target domains.}
   \label{figure3}
       \vspace{-0.15in}
\end{figure}
\subsection{Implementation Details}
We train  LLM4GTA on 4 A100 80G GPUs using AdamW optimizer with learning rate 1e-4 and cosine scheduling. The architecture employs SBERT as text encoder and GraphGPS as graph encoder, with a projection layer aligning representations to shared dimensionality. Following our methodology, we set gap threshold $\theta$ to 0.10 for citation networks and 0.12 for social networks, monitoring relative change $\delta$ in negative sample similarity as defined in Equation 8 to trigger early stopping. Fusion weight $\lambda$ is set to 0.8 and temperature $\tau$ to 0.2 across experiments. Training runs up to 140 epochs with batch size 256, though gap monitoring typically terminates between epochs 100-120 to preserve representation quality. We compute $\text{sim}_{pos}$ and $\text{sim}_{neg}$ every epoch on validation sets, applying exponential moving average with decay 0.9 for smoothing. Few-shot results are averaged over 5 random seeds. The framework adds less than 5\% overhead compared to standard contrastive training.


\subsection{Performance Comparison}
We compare LLM4GTA against 15 baselines across node classification in Table~\ref{table1} and Table~\ref{table3}, graph classification in Table~\ref{table2}. Our method achieves substantial improvements, surpassing the strongest baseline GCOPE by 16.75\% on Cora and 10.07\% on Photo for node classification, with gains reaching 25.8\% on LastFM social networks. Traditional GNN methods suffer from limited transferability without pre-trained representations, while recent LLM-based approaches like LLaGA and TEA-GLM remain constrained by naive alignment that disrupts geometric properties essential for transfer learning. Our framework addresses this through controlled gap preservation, maintaining structural inductive biases from graph encoders and semantic prior knowledge from language models. The dual-space fusion mechanism combines cross-modal semantic understanding with uni-modal discriminative power, enabling the model to leverage textual knowledge for zero-shot reasoning while utilizing graph-specific patterns for accurate classification. Consistent improvements across citation networks, e-commerce graphs, and social platforms validate that appropriate representation gaps protect domain-invariant knowledge, enabling robust transfer without catastrophic forgetting.
\begin{table*}[tbp]
    \centering
    \small
    \addtolength{\tabcolsep}{-0.14mm}
    \caption{Model ablation study on key components of \model.}
    \label{table4}%
    \begin{tabular}{c|ccc|ccc|ccc}
    \toprule
\multirow{2}*{Methods}
    & Gap & Graph & Dual-Space &\multicolumn{3}{c|}{Target domain for node classification} &\multicolumn{3}{c}{Target domain for graph classification}\\
    & Preservation & Classifier & Fusion & Cora & Photo & Facebook & Cora & Photo & Facebook\\
    \midrule
    \method{Variant 1}
    & $\times$ & $\times$ & $\times$ 
    &34.28 $\pm$ 14.35 & 47.82 $\pm$ 11.26 & 33.91 $\pm$ 10.78
    & 43.17 $\pm$ 15.29 & 50.68 $\pm$ 13.84 & 36.52 $\pm$ 11.93
    \\
    \method{Variant 2}
    & $\times$ & $\times$ & $\checkmark$ 
    &38.45 $\pm$ 13.67  & 54.87 $\pm$ 10.85  & 37.64 $\pm$ 12.14 
    &43.28 $\pm$ 15.76  & 55.93 $\pm$ 13.28  & 40.15 $\pm$ 12.67    \\ 
    \method{Variant 3} 
    & $\checkmark$ & $\times$ & $\times$
    & 42.13 $\pm$ 12.58  & 54.35 $\pm$ 11.89  & 38.94 $\pm$ 10.12 
    & 50.62 $\pm$ 14.17  & 56.27 $\pm$ 13.95  & 41.38 $\pm$ 11.42    \\  
    \method{Variant 4}
    & $\checkmark$ & $\checkmark$ & $\times$ 
    &43.82 $\pm$ 13.94  & 55.43 $\pm$ 11.67  & 38.21 $\pm$ 10.56
    &52.18 $\pm$ 16.25  & 56.04 $\pm$ 14.12  & 41.02 $\pm$ 11.78    \\ 
          \rowcolor[HTML]{E1EAFF}
    \method{\model}
    & $\checkmark$ & $\checkmark$ & $\checkmark$
    &\textbf{49.23} $\pm$ 14.56 &
    \textbf{60.48} $\pm$ 10.34 &
    \textbf{43.98} $\pm$ 10.87 &
    \textbf{57.12} $\pm$ 15.83 & \textbf{60.52} $\pm$ 13.45&
    \textbf{45.08} $\pm$ 11.26 \\
    \bottomrule
    \end{tabular}
\end{table*}

\subsection{Transfer Ability}
\begin{figure}[tb!]
    \centering
    \includegraphics[width=\columnwidth]{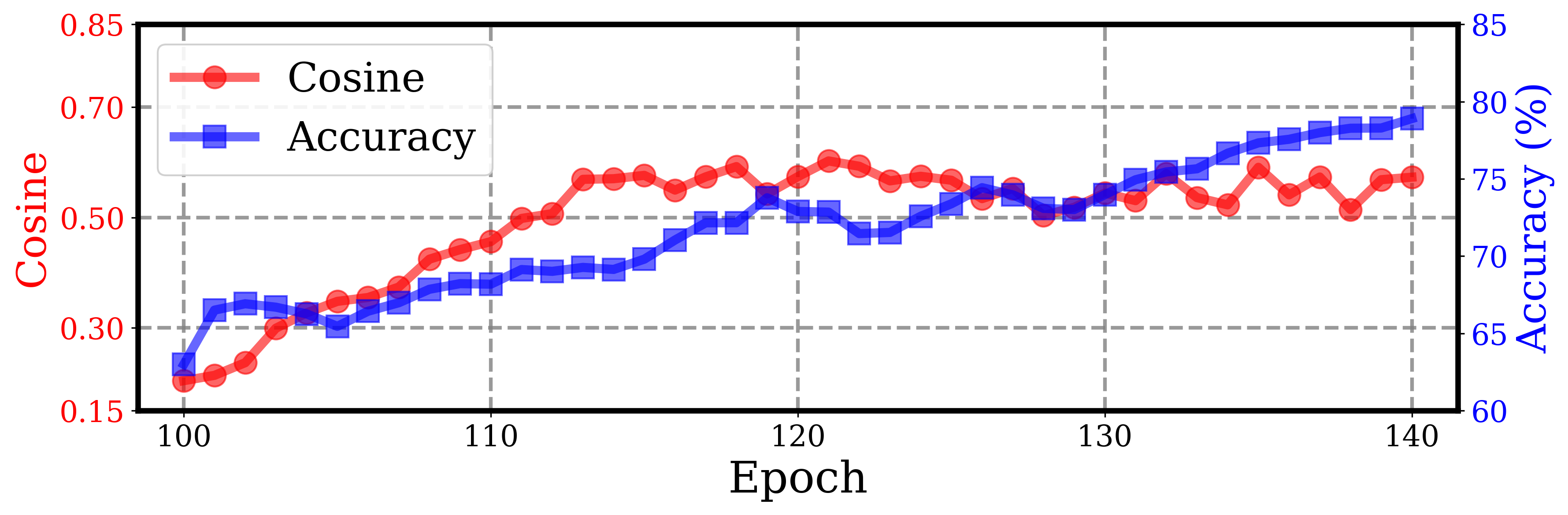}
    \caption{ LLM4GTA training curves showing stable gap preservation with consistent accuracy gains, validating our gap-aware alignment approach.}
    \label{figure4}
        \vspace{-0.15in}
\end{figure}

\begin{figure}[t]
    \centering
    \includegraphics[width=\columnwidth]{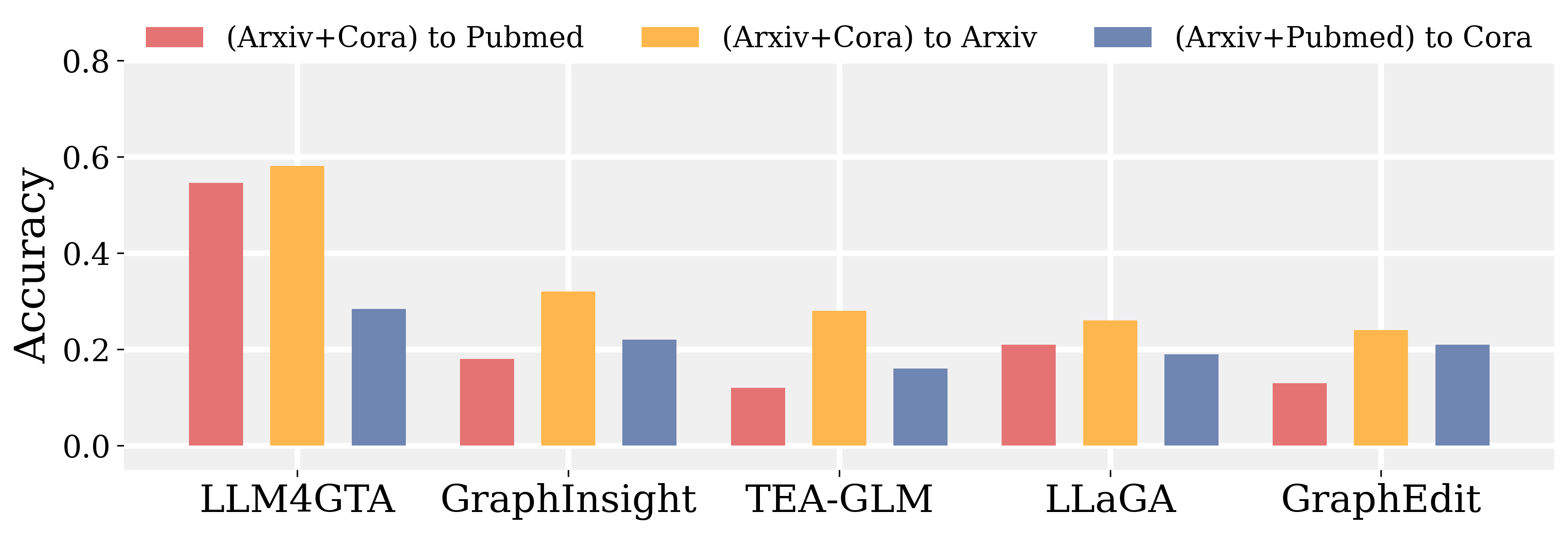}
\caption{Generalization performance of LLM4GTA on zero-shot node classification across multi-dataset transfer settings.}
    
    \label{figure5}
    \vspace{-0.15in}
\end{figure}

Figure~\ref{figure3} demonstrates that LLM4GTA maintains strong performance under data scarcity, with Cora accuracy transitioning from 70\% at 10-shot to 52\% at 1-shot and 45\% at zero-shot, compared to LLaGA's 28\% zero-shot performance. This sample efficiency arises from preventing over-alignment that would damage pre-trained representations. Figure~\ref{figure5} shows superior cross-dataset generalization, achieving 58\% accuracy when transferring from Arxiv and Cora to Pubmed versus GraphInsight's 19\%. The training dynamics in Figure~\ref{figure4} reveal crucial insights where cosine similarity rises moderately from 0.20 to 0.56 while accuracy improves from 65\% to 78\%. This decoupling validates that moderate gaps around 0.56 outperform both tight alignment above 0.75 causing structure collapse and loose coupling below 0.40 losing semantic correspondence. Our gap monitoring prevents the similarity-performance divergence in baselines, where aggressive alignment distorts pre-trained spaces.

\subsection{Hyperparameter Analysis}
\begin{figure}[t]
    \centering
    \includegraphics[width=\columnwidth]{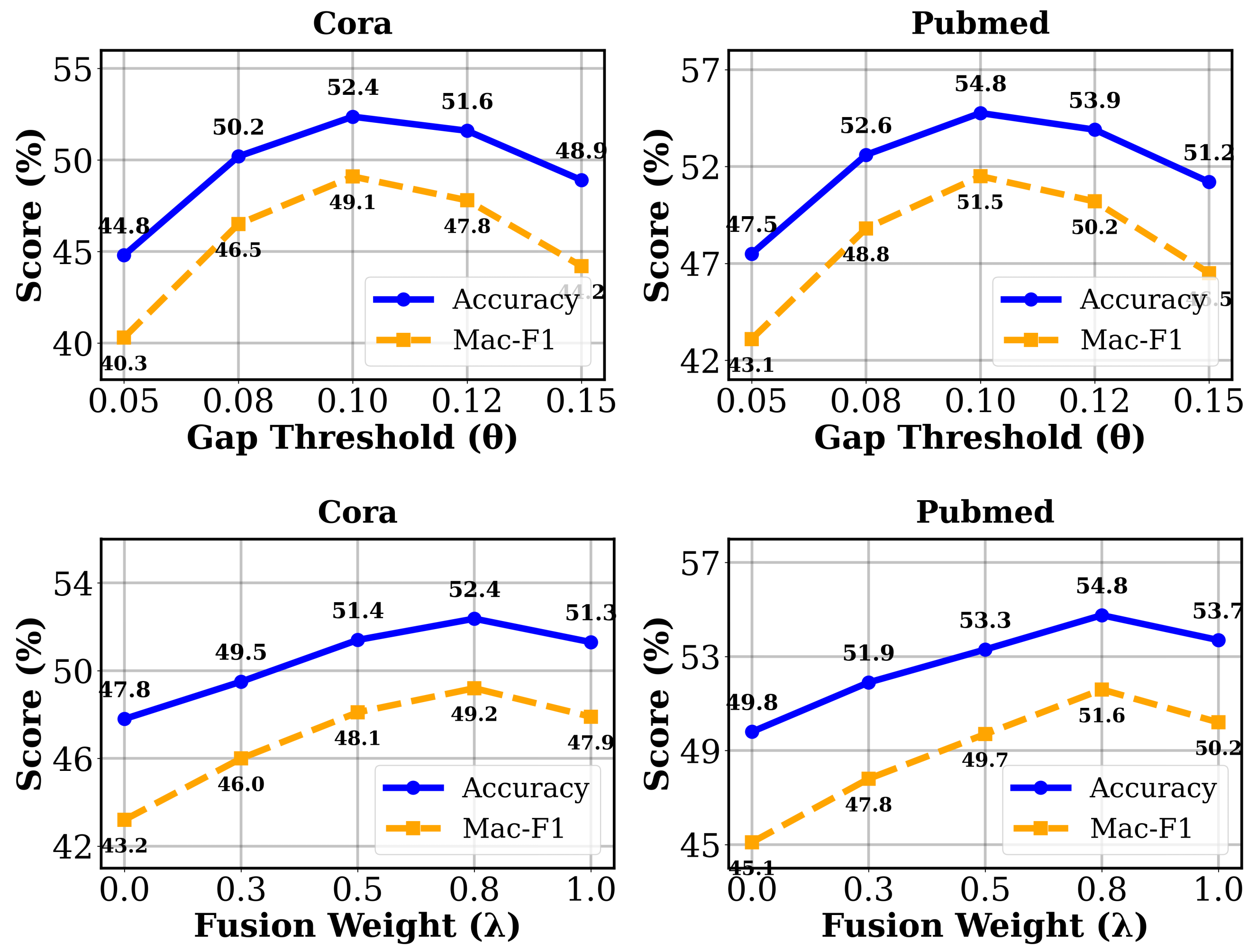}
\caption{Hyperparameter sensitivity analysis on one-shot node 
classification. }
    \label{figure6}
    \vspace{-0.15in}
\end{figure}

Figure~\ref{figure6} reveals that gap threshold $\theta$ exhibits an inverted-U relationship with performance. On Cora, accuracy peaks at 52.4\% when $\theta$ equals 0.10, declining to 46.1\% at 0.05 due to premature stopping and to 48.4\% at 0.15 from excessive alignment. This pattern validates that moderate gaps preserve both topological inductive biases and semantic structures. The fusion weight $\lambda$ demonstrates dual-space compensation necessity, where pure cross-modal prediction at zero weight achieves only 47.8\% on Cora versus 52.4\% at optimal $\lambda$ of 0.8, representing 9.6\% improvement. However, excessive fusion at 1.0 reduces performance to 51.2\% by over-relying on uni-modal classification and sacrificing cross-modal generalization benefits. The synergy between parameters illuminates our design philosophy where $\theta$ maintains geometric integrity while $\lambda$ compensates for discriminative limitations, jointly achieving robust transfer across diverse domains.

\subsection{Ablation Study}
Table~\ref{table4} systematically validates component contributions through progressive integration. The baseline without our components achieves 34.28\% on Cora node classification and 43.17\% on graph classification. Adding dual-space fusion alone improves to 38.45\%, providing 4.17\% gain but remaining limited. Introducing gap preservation yields substantial improvements to 42.13\% and 50.62\% respectively, representing 7.85\% and 7.45\% gains by preventing structure collapse. Combining gap preservation with graph classifiers achieves 43.82\% and 52.18\%, demonstrating complementary benefits. The complete framework reaches 49.23\% and 57.12\%, representing 43.6\% and 32.3\% relative improvements over baseline. The hierarchy reveals clear design rationale where gap preservation prevents manifold collapse, graph classifiers overcome dimensional constraints of text-only classification, and fusion balances semantic consistency with discriminative precision. This architecture collectively enables robust zero-shot and few-shot transfer by protecting domain-invariant knowledge while maintaining task-specific adaptability.

\section{Conclusion}
This paper reveals that representation gaps preserve domain-invariant knowledge essential for generalization, while over-alignment causes structure collapse. We propose LLM4GTA with adaptive gap preservation preventing collapse and intra-graph-space compensation enabling dual-space prediction fusion. Experiments validate our approach across zero-shot and few-shot settings, establishing that appropriate separation enhances transfer performance while preserving specialized knowledge structures.

\bibliographystyle{ACM-Reference-Format}
\bibliography{sample-base}

\end{document}